\documentclass[prl,twocolumn]{revtex4}
\usepackage{graphicx}
\usepackage{dcolumn}
\usepackage{amsmath}
\usepackage{amssymb}
\usepackage{bm}
\begin{document}
\newcommand{\be}{\begin{equation}}
\newcommand{\ee}{\end{equation}}
\newcommand{\vk}{\mathbf{k}}
\newcommand{\vq}{\mathbf{q}}
\newcommand{\vp}{\mathbf{p}}
\newcommand{\vx}{\mathbf{x}}
\newcommand{\vy}{\mathbf{y}}
\newcommand{\vz}{\mathbf{z}}
\newcommand{\vw}{\mathbf{w}}
\newcommand{\half}{\frac{1}{2}}
\newcommand{\la}{\langle}
\newcommand{\ra}{\rangle}
\newcommand{\p}{\partial}
\newcommand{\vn}{\hat{\mathbf{n}}}
\newcommand{\e}{\epsilon}
\title{ On two descriptions of disordered phase of 2D quantum  antiferromagnets: monopole plasma and Kalb-Ramond fields}
\author{Hyun C. Lee}
\email{ hyunlee@sogang.ac.kr}
\affiliation{Department of Physics and Basic Science Research Institute, 
Sogang University,Seoul, 121-742, Korea}
\date{\today}
\begin{abstract}
We have studied a system of light bosonic spinons interacting with compact gauge fields.
By the generalization of the works by Polyakov on compact gauge fields the system is mapped to monopole plasma model and 
a model of open surfaces coupled to antisymmetric Kalb-Ramond gauge fields.
The monopole correlation function in the presence of light spinon is computed based on these two models.
\end{abstract}
\maketitle
The quantum critical phenomena and the associated quantum phase transitions are one of the most 
outstanding problems of current physics \cite{ColemanNature,thebook}. 
These problems have been addressed mostly in the context of quantum magnetism and heavy fermion systems \cite{ColemanNature}.
The standard framework for the understanding of general critical phenomena is the Landau-Ginzburg-Wilson(LGW) theory\cite{lg,wilson}.
In LGW theory,  critical phenomena are described in terms of the order parameters characterizing either side of transition.

Recently a new framework of quantum criticality beyond LGW theory was proposed \cite{senthil1,senthil2}.
This framework predicts  degrees of freedom emerging at critical point 
which can not be expressed by order parameters of phases surrounding the critical point.
A very rough picture is that the emerging degrees of freedom are \textit{confined} away from critical point due to strong 
\textit{singular} ( or topologically nontrivial) gauge field fluctuations, 
while at the critical point the singular fluctuations become \textit{irrelevant} (in the sense of renormalization group)
 liberating  the confined degrees of freedom. As quarks cannot be described by mesons or baryons at low energy, the emerging deconfined 
degrees of freedom cannot be described by order parameters of either phase of critical point.

The key step of the above scenario is to show the irrelevancy of singular gauge field fluctuation at critical point.
This point has also been investigated intensively in the context of  massless (2+1)-dimensional quantum electrodynamics 
along with some controversies \cite{Hermele,Kleinert,Herbut}.
In (2+1)- dimensional system the singular gauge field fluctuations can be identified with  monopole field configurations 
\cite{polya0,polya1}. The monopoles can exist only if the associated U(1) gauge field is an angular variable ( in other words, 
has its value on cirle $S^1$ ), and such gauge field is called \textit{compact}. Ordinary U(1) gauge field has its value on $R^1$ and
is called non-compact. Polyakov showed that the \textit{static and heavy} charges interacting via compact gauge fields
have linearly rising interaction potential energy($V(R) \propto R$), thus they are in confinement \cite{polya1}.
The potential energy $V(R)$ is the energy of bundle of electric flux line between two charges, so it is naturally proportional to
the length of flux line.
The result by Polyakov \cite{polya1} cannot be directly applied to the problem of quantum criticality since the matter fields with
gauge charges are \text{dynamic and light} in the vicinity of quantum critical point.

In this paper we study the descriptions of the disordered phase two-dimensional quauntum antiferromagnets where  the interaction
between  the light bosonic matter degrees of freeom and  the compact gauge field (especially monopole configurations)
 can be expressed  in a very illuminating form. This formulation enables us to compute the correlation functions of monopole 
 operators in the presence of light spinon.

We start with the $O(3)$ nonlinear sigma model  formulation of spin-$S$ two-dimensional quantum antiferromagnets \cite{thebook}.
\begin{align}
\label{nlsm}
S_n &= S_0 + S_B \cr
S_0 &= \frac{1}{2 g} \, \int_0^\beta d \tau \int d^2 r \,\left [ (\p_\tau \vn)^2 + c^2 (\partial_{\vec{r}} \vn)^2 \right ] \cr
S_B &=i S  \sum_{\vec{r}}\,\e_{\vec{r}}\,\int_0^\beta d \tau  \int_0^1 du \, \vn \cdot \p_\tau \vn \times \p_u \vn,
\end{align}
where $\vn=\vn(\tau,\vec{r})$ is a three component N\'eel order parameter and 
$\e_{\vec{r}} = (-1)^{r_x + r_y}$. $S_B$ is the Berry phase term.
$\vec{r}$ is to be understood as
both  continuum and lattice coordinates. 
The large $N$ expansion study of $S_0$ shows that the N\'eel ground state is realized for
$g^2 < g^2_{\text{cr}}$, while disordered massive uniform quantum paramagnets obtains for the strong coupling regime 
$g^2 > g^2_{\text{cr}}$
\cite{thebook}.
Haldane showed that the Berry phase term is non-vanishing only for the monopole configuration of N\'eel order parameter
\cite{Haldane}.

The monopoles can be described by U(1) gauge field in the 
$\text{CP}^1$ representation of Eq. (\ref{nlsm}) \cite{thebook,SachdevJalabert}.
\begin{align}
\vn & = Z^\dag \boldsymbol{\sigma} Z, \quad \sum_{\alpha=1}^2 \vert Z_\alpha \vert^2=1 \cr
A_{\mu} &= - \frac{i}{2} \sum_\alpha\, \Big[Z^*_\alpha \p_\mu Z_\alpha -\text{C.C} \Big],\cr
F_{\mu \nu} &= \half \, \vn \cdot \p_\mu \vn\times \p_\nu \vn. 
\end{align}
$Z$ is called bosonic spinon in the context of quantum antiferromagnetism.
The compact nature of gauge field $A_\mu$ is manifest since it is the fiber of Hopf bundle
$S^3 (Z)  \to S^2 (\vn )$.
The monopole of strength $p$ is characterized by
\be
p = \frac{1}{2\pi}\, \int_{\Sigma} F, \quad F = \half F_{\mu \nu} dx^\mu \wedge d x^\nu,
\ee
where $\Sigma$ is a closed surface enclosing the monopole.
Read and Sachdev showed that the Berry phase term can be reduced to \cite{sn1,sn2}
\be
\label{berry2}
S_B = -i (2 S)\, \frac{\pi}{2} \sum_s \,\zeta_s\, p_s,
\ee
where $s$ denotes the site of dual lattice and $p_s$ is the strength of monopole located at $s$.
$\zeta_s=0,1,2,3$ is an integer valued field designating sublattice structure of dual lattice.
We note that for $p_s = 4$, $e^{-S_B}=1$, so that the Berry phase term disappears.

The $\text{CP}^1$ model can be  naturally generalized to $\text{CP}^N$ model and it  possesses a critical point.
Relaxing the constraint $ \sum_{\alpha=1}^N \vert Z_\alpha \vert^2 =1$ by mass term $m_Z$,
we obtain a model to be investigated in this paper.
\begin{align}
\label{action}
S & = S_Z + S_A + S_B \cr
S_Z &= \int d^3 x   \left(  \vert (\p_\mu - i A_\mu ) Z  \vert^2 + m_Z^2 \vert Z \vert^2 \right ) \cr
S_A & = \frac{1}{4 e^2}\,\int d^3 x \,F_{\mu \nu} F^{\mu \nu},
\end{align}
where $S_B$ is given by Eq. (\ref{berry2}). $e^2$ (which has dimension of mass) will be treated as phenomenological parameter
independent from $m_Z$.
This model is a slight modification of the one considered by Sachdev, Jalabert, and Park \cite{SachdevJalabert,SachdevPark}.
In their works the singular monopole gauge fields are expressed  separately in  action. 
\textit{ We will assume that $m_Z$ is light enough, so that only the monopoles with strength $p=4 \,\text{mod}\, 4$
 survive in the functional
integral}. If  $m_Z$ is large enough the matter $Z$ fields can be simply dropped in low energy sector. 
The monopole summation (see below for  details ) can be done with Berry phase taken into account,
and it leads to 
$\cos[ \varphi_s - \pi S \zeta_s ]$. The sine-Gordon interaction pins the $\varphi_s$, thus depending on values of $S$
spin-Peierls states with broken lattice translational symmetry obtain \cite{sn2}.

We can formally integrate out the spinon field $Z$ exactly from Eq. (\ref{action}).
It contributes
$$\delta S_Z = N \,\mathrm{Tr} \ln \Big[ (-i \p_\mu - A_\mu)^2 + m_Z^2 \Big ]$$
to the effective action of gauge fields.
Next we employ the well-known path integral representation of logarithm of an operator 
\begin{align}
&\mathrm{Tr} \ln \Big[ (-i \p - A)^2 + m_Z^2 \Big ] = \half \int_{\epsilon}^{\infty} \frac{d \tau}{\tau} 
\mathrm{Tr} e^{-\frac{\tau}{2} [(-i \p - A)^2 + m_Z^2]} \cr
&=\half \int_{\epsilon}^{\infty} \frac{d \tau}{\tau} e^{-\tau m_Z^2/2}\,\int_{X(0)=X(\tau)} D[X_\mu(t)] \cr
&\times \exp\Big[ -\half\int_0^\tau dt \dot{X}_\mu^2 + i \oint_C A(X^\mu) \Big] \cr
&\equiv \sum_C'\, W(C), \quad W(C) = e^{i \oint_C A(X)}.
\end{align}
$C$ denotes the \textit{closed} trajectory of the $Z$-particle. $W(C)$ is the Wilson loop \cite{wilsonloop}.
$\e$ is a short time cut-off.

Then the partition function reads
\begin{align}
\label{partition}
Z &= \int D[A]\,e^{-S_A - N \sum_C'\, W(C) } \cr
&=\sum_{m=0}^\infty\,\frac{(-)^m N^m}{m!}\,\sum_{C_1}' \cdots \sum_{C_m}' \, \la W(C_1) \cdots W(C_m) \ra.
\end{align}
The average of Wilson loop is given by
\begin{align}
\la W(C_1) \cdots W(C_m) \ra &=\int D[A]\,e^{-S[A]}\, e^{i \sum_j \oint_{C_j} A} \cr
&=\int D[A]\,e^{-S[A]}\, e^{i \sum_j \int_{\Sigma_j} F},
\end{align}
where $\Sigma_j$ is an open surface whose boundary is $C_j$.
The gauge field can be decomposed into non-compact  and monopole part \cite{polya0}.
Accordingly the averages of Wilson loops are factorized
\begin{align}
\la W(C_1) \cdots W(C_m) \ra &= \la W(C_1) \cdots W(C_m) \ra_0  \cr
&\times\la W(C_1) \cdots W(C_m) \ra_M.
\end{align}
At low energy the monopole contributions dominate and we focus on the $\la W(C_1) \cdots W(C_m) \ra_M$.

The partition function of monopole plasma is given by \cite{polya0,polya1}
\be
Z_0 = \sum_{N=0[\{q_a\}}^\infty \,\frac{z^N}{N!}\,\int \prod_{j=1}^N d \vx_a\,
\exp\Big[- \frac{\pi}{2 e^2}\, \sum_{a \neq b}\, \frac{q_a q_b}{\vert \vx_a -\vx_b \vert} \Big ].
\ee
The fugacity is 
\be
z = \Lambda^3 \, \sqrt{\frac{ \Lambda}{e^2}}\, e^{-c_1 \Lambda/e^2},
\ee
where $\Lambda$ is a high energy cutoff and $c_1$ is a numerical constant of order unity.
The summation over monopoles can be done with Wilson loops inserted, and it results in
\begin{align}
\label{monopolesum}
&\la W(C_1) \cdots W(C_m) \ra_M = \frac{1}{Z_0}\,\int D[\varphi]\, \cr
&\times \exp\Big[  - \frac{e^2}{8\pi^2} \int d^3 x (\nabla \varphi)^2 +2 z \int d^3 x \cos[p ( \varphi + \eta)] \Big] \cr
& Z_0 =\int D[\varphi]\,\exp\Big[  - \frac{e^2}{8\pi^2} \int d^3 x (\nabla \varphi)^2 +2 z \int d^3 x \cos[p \varphi ] \Big],
\end{align}
where 
\be
\label{multipleloop}
\eta(\vx) = \sum_{\Sigma_j} \int  \frac{1}{2} \frac{( \vy_j -\vx) \cdot d \vec{a}(\vy_j)}{\vert \vy_j -\vx \vert^3}
\ee
is the source term of multiple Wilson loops. $\vy_j$ is the coordinates on the surface $\Sigma_j$.
Eq. (\ref{multipleloop}) is nothing but the solid anlge subtended by surface $\Sigma$ observed from the point $\vx$.
If particles carry nontrivial spinor, color, or flavor indices, the mutiple Wilson loops contain noncommuting matrices,
so that it cannot be expressed in a simple way as Eq. (\ref{monopolesum}).
$\varphi$ represents the photon in dual representation. As can be seen in $Z_0$ of Eq. (\ref{monopolesum}) $\varphi$ is pinned 
at the bottom of cosine potential and 
the photon becomes massive with exponentially small mass
\be
m_p \sim \Lambda \Big( \frac{\Lambda}{e^2} \Big)^{3/4} e^{-c_1 \Lambda/2 e^2}.
\ee
The partition function Eq. (\ref{partition}) and the average of Wilson loops of the form Eq. (\ref{monopolesum}) 
with appropriate operator insertions constitute a monopole plasma representation of light bosonic matter field coupled
to compact gauge fields.

Doing the monopole functional integral via steepest descent for a single \textit{large specified loop}  we find that
\be
\label{arealaw}
\la W(C) \ra  \sim \exp[ - \sigma S ], \quad \sigma \sim e^2 m_p,
\ee
where $S$ is the \textit{minimal area} bounding the trajectory $C$ \cite{polya1}.
From the viewpoint of Hamiltonian formulation Eq. (\ref{arealaw}) implies the linear potential, thus confinement.

In confining phase there exist electric field flux lines connecting two charges, and the time evolution of these lines generate 
open surfaces. Therefore, it is tempting to attempt to express the average of Wilson loop Eq. (\ref{monopolesum}) in terms of these 
surfaces. This has been done by Polyakov \cite{polya2}.
Consider a functional integral
\begin{align}
Z[\{ W(C) \} ]&=  \int D[\phi, B]\,e^{-\Gamma[\phi,B,\Sigma_C]} \cr
\Gamma[\phi,B,\Sigma_C]&= \int d^3 x \Big [ \frac{1}{4 e^2} B_{\mu \nu} B^{\mu \nu} + \frac{i}{2\pi} \phi \wedge d B \cr
&+(2z) ( 1- \cos p \phi) \Big ]  -i \int_{\Sigma_C} B.
\end{align}
$B_{\mu \nu}$ is an antisymmetric Kalb-Ramond gauge field \cite{Kalb} which naturally couples to surface just as vector potential
couples to point particle in a minimal way $\int d x^\mu A_{\mu}$.
$B=\half B_{\mu \nu} d x^\mu \wedge d x^\nu$ is a differential 2-form. $\Sigma_C$ is an open surface bounding closed path $C$.
The generalization to multiple surfaces is immediate. 
$$ -i  \int_{\Sigma_C} B  \to -i \sum_j \int_{\Sigma_{C_j}} B.$$

With $B_{\mu \nu}$ eliminated by solving equation of motion, it can be shown that
\be
\label{equivalence}
W_o(C) W_M(C)\propto Z[\{ W(C) \} ],\quad   \phi- \eta = \varphi.
\ee
This establishes the equivalence between monopole plasma approach and the Kalb-Ramond gaug field approach.
Now one can integrate over $\phi$ first to obtain effective action of $B_{\mu \nu}$, and the result still should be 
the same as the Wilson loop $W(C)$. In three dimension 
$$
d B = H dx^1 \wedge d x^2 \wedge d x^3,\;\; H= \half \epsilon^{\lambda \mu \nu} \p_\lambda B_{\mu \nu}.
$$
Eliminating $\phi$ by solving equation of motion of $\phi$ we obtain
\begin{align}
\label{kalb}
W(\{ C_j \}) &= \int D[B]\, e^{- S(B) } e^{ -i  \int_{\Sigma_{C_j}} B}, \cr
S(B) &= \int d^3 x \Big( \frac{1}{4 e^2} B_{\mu \nu} B^{\mu \nu} \cr
& +\frac{1}{2\pi p} \Big[ H \sin^{-1} \frac{H}{m_H^3}
- \sqrt{m_H^6 + H^2} \Big] \Big ),
\end{align}
where
\be
m_H = \Lambda \Big( \frac{\Lambda}{e^2} \Big )^{1/6} e^{-c_1 \Lambda/3 e^2}.
\ee
Polyakov has obtained a non-standard string theory by summing over branches of $\sin^{-1} \frac{H}{m_H^3}$ \cite{polya2}.
The partition function Eq.(\ref{partition}) and the averages of Wilson loop Eq.(\ref{kalb}) constitute a description
in terms of Kalb-Ramond antisymmetric gauge field.
Picking up a branch near $H=0$ and expanding in $H$ for large loops the effective action $S(B)$ becomes
\be
\label{quadratic}
S(B)= \int d^3 x \Big( \frac{1}{2 e^2} \vec{B} \cdot \vec{B}+ \frac{1}{4\pi p m_H^3} (\nabla \cdot \vec{B})^2 \Big ),
\ee
where
$B^i = \half \epsilon^{ijk} B_{jk}$.
The propagator is 
\be
\la B_i(\vx) B_j(\vy) \ra = \delta_{ij}\, \frac{p m_H^3}{2}\,\frac{e^{- m_p \vert \vx -\vy \vert}}{ \vert \vx -\vy \vert}.
\ee
Now the averages of Wilson loop can be computed from Eq.~(\ref{kalb})
\begin{align}
W(\{C_j \} ) &= \exp \Big[-\frac{p m_H^3}{4}\, \sum_{i,j} \int \int d \sigma^a(\vx_i) d\sigma^a(\vy_j) \cr
 &\times \frac{e^{- m_P \vert \vx_i -\vy_j \vert}}{4\pi \vert \vx_i -\vy_j \vert} \Big ].
\end{align}
For a very  large loop the main contributions will come from the region $ \vert \vx -\vy \vert \sim m_p^{-1}, \,
d \sigma^a \sim m_p^{-2}$.
The expectation value of single loop is then
 \be
 \label{minimal}
 \la W(C) \ra  \sim \exp\Big[ - m_p c_1 e^2 \int d^2 \xi \sqrt{g}  \Big], 
 \ee
 where
 \be
 g_{ab} = \p_a \vec{x} \cdot \p_b \vec{x}
 \ee
is the induced metric. 
Thus the minimal surface with prescribed boundary will give the dominant contribution to the Wilson loop (see Eq.~(\ref{arealaw})).
 
Now let us discuss the correlation function of monopole operator $\la e^{i p \varphi(\vz)} e^{-ip \varphi(\vw)} \ra$ with
$\vert \vz -\vw \vert \gg m_p^{-1}$.
 
We note that the monopole operator can be identified with the valence bond operator \cite{sn2,sn3}.
\be
e^{i \varphi(\vec{r})} \sim (-1)^{r_x + r_y} \vec{S}(\vec{r}) \cdot \vec{S}(\vec{r}+\vec{\mu}).
\ee
Since the monopole operator acts trivially on non-compact sector the correlation function can be expressed by
\begin{align}
\label{correlator}
\la e^{i p \varphi(\vz)} e^{-ip \varphi(\vw)} \ra &= 
\frac{1}{Z_M}\,\sum_m \frac{(-1)^m N^m}{m!}  \Sigma_{C_1}' \cdots \Sigma_{C_m}' \cr
&\times \la e^{i p \varphi(\vz)} e^{-ip \varphi(\vw)} \prod_j \la W(C_j) \ra_M,
\end{align}
where $Z_M$ is the partition function in monopole sector
\be
\label{correlator1}
Z_M = 
\sum_m \frac{(-1)^m N^m}{m!}  \Sigma_{C_1}' \cdots \Sigma_{C_m}' 
\la \prod_j W(C_j) \ra_M.
\ee
Let us calculate $m=0$ and $m=1$ contributions to the correlation functions.
The spinon is absent in $m=0$ sector, and the correlation function is essentially determined by pinning and massive photons.
\be
\la e^{i p \varphi(\vz)} e^{-ip \varphi(\vw)} \ra_{m=0} \sim  1 - \frac{e^{-m_p \vert \vz -\vw \vert}}{ \vert \vz -\vw \vert}.
\ee
In the computation of $m=1$ contribution it is essential to include the $m=1$ part of $Z_M$.
\begin{align}
\label{normalization}
Z_M(m=1) & = -N \sum_C' \la W(C) \ra_M \cr
&\approx -N \sum_C' \exp[ - \sigma S(C) ] \cr
&\approx -N \int_\e^\infty \frac{d \tau}{\tau} \tau^{-D/2} e^{-m_Z^2 \tau/2- \sigma S(C)}
\end{align}
where the second line is from Eq. (\ref{arealaw}) and the summation over closed paths can be done by following Chap.9 of 
Ref.\cite{polya0}. In the above $m_Z$ is to be understood as a mass in continuum theory. $D$ is a space-time dimension.
For spinless particle the length of typical \textit{dynamical} path $\tau$ is 
the order of area $S(C)$ (like Brownian motion) \cite{polya3}. (For fermions $\tau \sim \sqrt{S(C)}$)

Next we compute the $m=1$ part of the numerator of Eq.(\ref{correlator}).
\begin{align}
-N \frac{1}{Z_0}\,\int D[\varphi]\,\exp\Big[-\int \frac{e^2}{8\pi^2} (\nabla \varphi)^2 \cr
+2z \cos(p (\varphi +\eta))+i p \varphi(\vw) -i p \varphi(\vz) \Big ].
\end{align}
For a large loop the functional integral can be computed in a saddle point approximation.($\phi = \varphi + \eta$)    
\be
\label{saddle}
\frac{e^2}{4\pi} \nabla^2 (\phi - \eta) = 2 p z \sin p \phi + ip \delta(\vx-\vz) -ip \delta(\vx-\vw).
\ee
$\nabla^2 \eta$ is the source term from the spinon loop.

The investigation of Eq. (\ref{saddle}) reveals that when $\vz, \vw$ are far
away from surface bounding boson loop, the correlation  is rather small.
$$  \exp \left[-\sigma S(C) + \frac{e^{-m_p |\vz-\vw|}}{|\vz-\vw|} \right].$$

Main contributions come from the cases where the monopole operators are located on the surface $\Sigma_C$.
In this case the local perturbation of $\varphi$ field can be relaxed by 
soft transverse modes of surface whoe energy is of the order
\be
E \sim \frac{1}{R^2},
\ee
where $R$ is the size of the closed path \cite{polya2}. 
This is the way how matther fields interact with monopoles.
Thus for the surfaces on which the monopole operators reside the correlation becomes
$$   \exp \left[-\sigma S + \frac{e^{- |\vz-\vw|/e^2 R^2}}{|\vz-\vw|} \right].$$
Taking into the normalizations Eq. (\ref{normalization})
we obtain
\begin{align}
& \la e^{i p \varphi(\vz)} e^{-ip \varphi(\vw)} \ra \sim 
 1 - \sum_C' e^{-\sigma S} \Big( \exp \left[ \frac{e^{- |\vz-\vw|/S}}{|\vz-\vw|} \right] -1 \Big ) \cr
&=1-  \frac{1}{\vert \vw -\vz \vert }\,\sum_C' e^{-\sigma S -\vert \vw -\vz \vert / S},
\end{align}
where $S(C) \sim R^2$ is used.
Employing the scaling $\tau \sim S(C)$ 
\begin{align}
\label{result1}
&\sum_C' e^{-\sigma S -\vert \vw -\vz \vert / S} = \int_\e^\infty \frac{d \tau}{\tau} \tau^{-\gamma/2}\, \cr
&\times\exp \Big[ - \frac{m_Z^2 \tau}{2} - \sigma \tau - \frac{\vert \vw -\vz \vert}{  \tau} \Big ] \cr
&\approx \left(\frac{\vert \vw -\vz \vert}{m_z^2+\sigma} \right)^{-\gamma/4}\,K_1( \sqrt{\vert \vw -\vz \vert(m_z^2/2+\sigma)}),
\end{align}
where $K_1$ is the modified Bessel function.
From Eq. (\ref{result1}) it is clear the massive boson suppressess the monopole fluctuation.
The exponent $\gamma$ cannot be put to 3 since only the subclass of surfaces contribute to Eq. (\ref{result1}).
Since the addition of one more point determines a plane $\gamma$ may be approximated to be 1.
The monopole correlation function is significantly influenced only in the  limit where the \textit{renormalized spinon mass}
 $m_Z^2 + 2\sigma$ 
becomes exceedingly small. In this limit 
\be
\label{main}
\la e^{i p \varphi(\vz)} e^{-ip \varphi(\vw)} \ra \sim 1 - \frac{\text{const.}}{\vert \vw - \vz \vert^{\gamma/4+3/2}}.
\ee
The result Eq. (\ref{main}) clearly shows that light spinons tend to prevent the proliferation of monopoles.
The contributions from $m \ge 2$ terms involve 
 various configurations of surfaces, such as foldings and  intersecting surfaces. 
This problems are left for future investigations.

In summary we have investigated two descriptions of the system of bosonic spinons interacting with compact gauge field.
These descriptions provide rather clear picture on the interaction of light spinons with monopole excitation of gauge field.
Monopole correlation functions is computed up to the first order of spinon loop and it turns out that the soft modes of 
the surface bounding spinon trajectory plays a crucial role.

The author is thankful to Chanju Kim for useful comments.


\end{document}